\documentclass[sigconf]{acmart}

\acmConference[ASE 2026]{The 41st IEEE/ACM International Conference on Automated Software Engineering}{October 12–16, 2026}{Munich, Germany}
\setcopyright{none}
\renewcommand\footnotetextcopyrightpermission[1]{}

\usepackage{xspace}
\usepackage[utf8]{inputenc}
\usepackage[most]{tcolorbox}
\usepackage[linesnumbered,ruled,vlined]{algorithm2e}
\usepackage{amsmath,amsfonts}
\usepackage{enumerate}
\usepackage{enumitem}
\usepackage{multirow}
\usepackage{adjustbox}
\usepackage{graphicx}
\usepackage{threeparttable}
\usepackage{booktabs}   
\usepackage{tabularx}   
\usepackage{ragged2e}   
\usepackage[font=small,skip=2pt]{caption} 
\usepackage[normalem]{ulem}  
\usepackage[table]{xcolor}
\usepackage{makecell}
\usepackage{subcaption} %
\usepackage{pifont}   
\usepackage{makecell} 

\newcommand{\hm}{\textsc{HybridMonkey}\xspace}
\newcommand{\hd}{\textsc{HybridDroidbot}\xspace}
\newcommand{\llmd}{\textsc{LLMDroid}\xspace}
\newcommand{\au}{\textsc{Aurora}\xspace}
\newcommand{\vet}{\textsc{Vet}\xspace}
\newcommand{\fb}{\textsc{Fastbot}\xspace}
\newcommand{\mk}{\textsc{Monkey}\xspace}
\newcommand{\rmk}{\textsc{Monkey*}\xspace}
\newcommand{\rd}{\textsc{Droidbot-random}\xspace}
\newcommand{\db}{\textsc{Droidbot}\xspace}

\newcommand{\ui}{\textsc{Uiautomation}\xspace}
\newcommand{\utwo}{\textsc{Uiautomator2}\xspace}

\newcommand{\gptd}{\textsc{GPTDroid}\xspace}
\newcommand{\themis}{\textsc{Themis}\xspace}

\newcommand{\totalBug}{75\xspace}
\newcommand{\reportBug}{34\xspace}
\newcommand{\fixAndConfirm}{\textsc{26}\xspace}
\newcommand{\fixBug}{18\xspace}
\newcommand{\confirmBug}{8\xspace}

\newcommand{\newBug}{34\xspace}

\newif\ifshowannotations
\showannotationstrue   

\ifshowannotations
  \usepackage{xcolor}
  \newcommand{\wan}[1]{{\color{cyan!70!blue}{(Wan: #1)}}}       
  \newcommand{\fixed}[1]{\textcolor{gray}{[Fixed: #1]}} 
  \newcommand{\deleted}[1]{\textcolor{red}{\sout{#1}}}
\else
  \newcommand{\wan}[1]{}            
  \newcommand{\fixed}[1]{}       
  \newcommand{\deleted}[1]{}
\fi

\newcolumntype{Y}{>{\RaggedRight\arraybackslash}X}

\definecolor{todocolor}{rgb}{0.9,0.1,0.1}

\newcommand{\eg}{\hbox{\emph{e.g.}}\xspace}
\newcommand{\ie}{\hbox{\emph{i.e.}}\xspace}

\usepackage{pythonhighlight}

\definecolor{codegreen}{rgb}{0,0.6,0}
\definecolor{codegray}{rgb}{0.5,0.5,0.5}
\definecolor{codepurple}{rgb}{0.58,0,0.82}
\definecolor{backcolour}{rgb}{0.97,0.97,0.95}
\definecolor{forestgreen}{rgb}{0.28,0.62,0.37}

\lstdefinestyle{mystyle}{
    backgroundcolor=\color{backcolour},   
    commentstyle=\color{codegray},
    keywordstyle=\color{codepurple},
    numberstyle=\tiny\color{codegray},
    stringstyle=\color{blue},
    basicstyle=\ttfamily\footnotesize,
    breakatwhitespace=false,         
    breaklines=true,                 
    captionpos=b,                    
    keepspaces=true,                 
    numbers=left,                    
    numbersep=5pt,                  
    showspaces=false,                
    showstringspaces=false,
    showtabs=false,                  
    tabsize=4,
}

\makeatletter
\newcommand{\removelatexerror}{\let\@latex@error\@gobble}
\makeatother

\begin{document}

\title{Improving Random Testing via LLM-powered UI Tarpit Escaping for Mobile Apps}

\author{Mengqian Xu}
\affiliation{%
  \institution{East China Normal University}
  \city{Shanghai}
  \country{China}
}
\affiliation{%
  \institution{Shanghai Innovation Institute}
  \city{Shanghai}
  \country{China}
}
\email{xmengqian@stu.ecnu.edu.cn}

\author{Yiheng Xiong}
\affiliation{%
  \institution{East China Normal University}
  \city{Shanghai}
  \country{China}
}
\email{yihengx98@gmail.com}

\author{Le Chang}
\affiliation{%
  \institution{East China Normal University}
  \city{Shanghai}
  \country{China}
}
\email{10225101547@stu.ecnu.edu.cn}

\author{Ting Su}
\affiliation{%
  \institution{East China Normal University}
  \city{Shanghai}
  \country{China}
}
\email{tsu@sei.ecnu.edu.cn}

\author{Chengcheng Wan}
\affiliation{%
  \institution{East China Normal University}
  \city{Shanghai}
  \country{China}
}
\affiliation{%
  \institution{Shanghai Innovation Institute}
  \city{Shanghai}
  \country{China}
}
\email{ccwan@sei.ecnu.edu.cn}

\author{Weikai Miao}
\affiliation{%
  \institution{East China Normal University}
  \city{Shanghai}
  \country{China}
}
\email{wkmiao@sei.ecnu.edu.cn}
\begin{abstract}
Random GUI testing is a widely-used technique for testing mobile apps. 
However, its effectiveness is limited by the notorious issue --- \emph{UI exploration tarpits},
where the exploration is trapped in local UI regions, thus impeding test coverage and bug discovery. 

In this experience paper, we introduce \textit{LLM-powered random GUI Testing}, a novel hybrid testing approach to mitigating UI tarpits during random testing. Our approach monitors UI similarity to identify tarpits and query LLMs to suggest promising events for escaping the encountered tarpits. We implement our approach on top of two different automated input generation (AIG) tools for mobile apps: (1) \hm upon \mk, a state-of-the-practice tool; and (2) \hd upon \db, a state-of-the-art tool.  We evaluated them on 12 popular, real-world apps. The results show that \hm and \hd outperform all baselines, achieving average coverage improvements of 54.8\% and 44.8\%, respectively, and detecting the highest number of unique crashes. In total, we found \totalBug unique bugs, including \newBug previously unknown bugs. To date, \fixAndConfirm bugs have been confirmed and fixed. We also applied \hm on \textit{WeChat}, a popular industrial app with billions of monthly active users. \hm achieved higher activity coverage and found more bugs than random testing.
\end{abstract}
    
\maketitle

\section{Introduction}

\label{sec:introduction}

Ensuring the reliability of mobile applications (apps) is critical for user retention. 
In practice, manual testing is prevalent~\cite{kochhar2015understanding,linares2017developers}, although it is usually small-scale, labor-intensive, and likely to miss bugs. To this end, a number of automated input generation (AIG) techniques, \eg, random, model-based, and learning-based testing, have been proposed in the past decade~\cite{monkey,machiry2013dynodroid,mao2016sapienz,li2017droidbot,lv2022fastbot2,choudhary2015automated}.

\begin{figure}[t]
    \vspace{0.3cm}
    \centering
    \includegraphics[height=4cm]{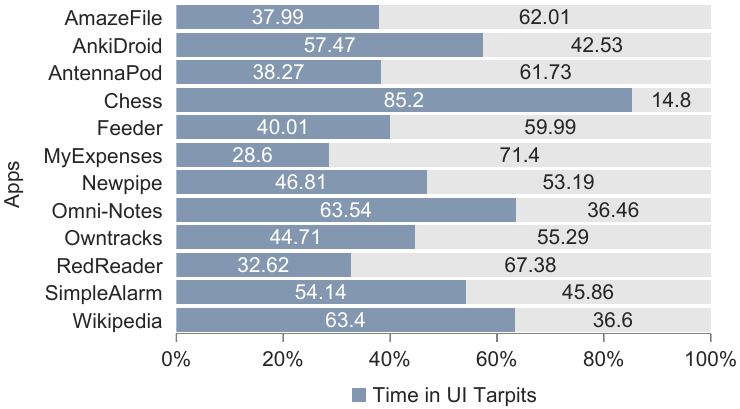}
    \caption{Proportion of time spent on UI tarpits on 12 real-world Android apps.}
    \label{fig:tarpit-distribution}
    \vspace{-0.5cm}
\end{figure}

Recent studies have shown that random testing is still one of the most effective UI testing techniques, often outperforming other sophisticated testing techniques, in both achieved code coverage and the number of found bugs~\cite{choudhary2015automated,patel2018effectiveness,wang2018empirical,zeng2016automated,mohammed2019empirical,lan2024navigating}.
This advantage stems from its simplicity and efficiency --- quickly generating a large number of events and reaching deep app states. 
Indeed, random testing is still one of the most widely-used testing techniques in the industrial setting~\cite{zeng2016automated,choudhary2015automated,patel2018effectiveness,wang2018empirical}.
However, random testing is likely to be trapped in \textit{UI exploration tarpits}~\cite{wang2021vet}, where it may get stuck in some local UI regions and fail to achieve fruitful exploration. One important reason is that random testing is semantics-oblivious. It is difficult for random testing to interpret the semantics and contexts of UI elements on the UI pages, thus leading to narrow exploration. Indeed, our preliminary study reveals that random testing wastes nearly 50\% of testing time in UI tarpits (see \S\ref{sec:pri_study}, Figure~\ref{fig:tarpit-distribution}). 

To our knowledge, \vet~\cite{wang2021vet} and \au~\cite{khan2024aurora} are the only two work that tackles UI exploration tarpits. 
However, these work have some major limitations. 
First, \vet   simply disables specific UI actions to avoid entering the region of UI tarpits, which likely leads to insufficient testing of the apps under test.
Second, although \au aims to overcome the shortcomings of \vet, it is limited to eight heuristic rules for specific UI patterns; thus, it struggles to generalize to unseen scenarios, especially given that the apps are diverse and frequently updated (discussed in \S\ref{aurora-tarpit}).

To tackle this problem, we introduce \emph{LLM-powered random GUI testing}, a novel hybrid testing approach to mitigating UI tarpits during random testing. Our \emph{key idea} is to interleave random testing with large language models (LLMs) guided exploration to escape UI tarpits. Specifically, we leverage LLMs to understand and provide guidance to escape the UI tarpits. 
This hybrid testing approach combines the strengths of random testing that \emph{achieves deep exploration and thus exhibits different app states}, and LLM-guided exploration that \emph{escapes UI tarpits and thus enables wide exploration}.

To achieve our idea, we monitor the transitions of UI pages to detect UI tarpits when performing random testing. 
Specifically, we use \emph{a number of consecutive and visually similar UI pages} as an intuitive yet common pattern to detect UI tarpits. Once a UI tarpit is detected, our approach switches from random testing to the LLM-guided exploration. 
During this stage, our approach invokes an LLM to analyze UI elements and execution history to infer events that are likely to escape the encountered UI tarpit. Once the tarpit has been successfully escaped, our approach resumes random testing.
These two stages are interleaved throughout the testing campaign until the time budget is exhausted. 
To further enhance efficiency, our approach caches the encountered UI tarpits and selectively reuses the events suggested by LLM in history.


We have realized our approach on two different AIG tools: (1) \hm upon official Android \mk~\cite{monkey}; and (2) \hd upon \db~\cite{li2017droidbot}, a popular academic AIG tool. We evaluated our tools against seven commonly-used and state-of-the-art baselines on 12 popular Android apps from Google Play. Results show that \hm and \hd consistently outperform all baselines in code coverage and bug detection.
Specifically, we achieve $2\mathrm{X}$ line and activity coverage of \au which tackles UI tarpits.
Compared to the best traditional baseline, our approach improves line, branch, and activity coverage by 17.9\%, 21.4\%, and 10.1\%, respectively. 
These improvements increase to 27.3\%, 39.5\%, and 45.6\% when compared to the best LLM-based baseline.
Under the same testing budget, \hm detects about $2\mathrm{X}$ more unique crashes than the best baseline, which clearly demonstrates the benefits of our approach.

To further assess the bug finding ability of our approach, we applied \hm to the latest versions of these 12 apps available at the time of our experiment (10 runs of 3-hour testing).
In total, \hm uncovered \totalBug unique crashes.
Among them, \newBug are previously unknown bugs.
The remaining crashes include 6 regressions and 35 known bugs that were independently found by \hm.
To date, \fixAndConfirm of the newly reported bugs have been confirmed or fixed by the developers, while the others remain under discussion.
In our approach, LLM-powered tarpit escaping achieves on average 72.9\% success rate.
Indeed, we observe that successful escapes typically lead to new code coverage within a short period of time.
Extended evaluations on two additional datasets (\S\ref{sec:discussion}) yield results which are consistent with our primary findings. A cost analysis reveals that our approach is the most cost-effective among the compared LLM-based tools. These findings highlight the practicability of our hybrid testing approach.

In summary, this paper has made the following contributions:
\begin{itemize}[leftmargin=*]
\item We propose a novel hybrid testing approach which interleaves random testing with LLM-guided exploration to escape UI exploration tarpits and thus improve testing effectiveness.
\item We instantiate our approach as two tools \hm and \hd by extending two existing AIG tools, demonstrating the applicability of our approach.
\item We conduct extensive experiments on 12 real-world Android apps against seven state-of-the-art baselines, showing our approach significantly improves code coverage and finds more bugs. 
\end{itemize}

\section{Observation and Illustrative Example}

\label{sec:motivation}

\subsection{Prevalence of UI Tarpits}
\label{sec:pri_study}

Despite the simplicity and efficiency of random testing, its effectiveness is often constrained by a pervasive phenomenon: UI tarpits. A UI tarpit occurs when a testing tool is trapped in a loop of visually similar screens that only support a small subset of app functionality.
In UI tarpits, a testing tool can only navigate within a narrow portion of the app. To assess the prevalence of UI tarpits in real-world settings, we conducted a preliminary study by applying random testing on 12 real-world Android apps.
Each app was tested for three hours, and we report the average results over three independent runs to ensure reliability. 
We capture a screenshot after each event and analyze the post-execution screenshot to identify tarpits.
In this preliminary study, a UI tarpit is identified when there are more than eight consecutive screenshots with high visual similarity (details in \S\ref{def:ui-tarpit}). 
We calculate the time wasted in the tarpits based on the timestamps of the captured screenshots. We compute the time elapsed between the first and last events of a tarpit.

The results are shown in Figure~\ref{fig:tarpit-distribution}. Most apps spent nearly half of the total testing time in UI tarpits. Notably, the test procedure of \textit{Chess} spent 85.20\% of its time trapped in UI tarpits. This demonstrates that the time wasted in these traps is significant.
Meanwhile, different tarpits may consume different amounts of time. \textit{Simple Alarm} encountered the most UI tarpits (182 occurrences), which consumed 54.14\% of the testing time. 
These findings motivate us to proactively recognize and escape UI tarpits during GUI testing. 



\subsection{An Illustrative Example}

\begin{figure}[t]
    \centering
    \includegraphics[width=\linewidth]{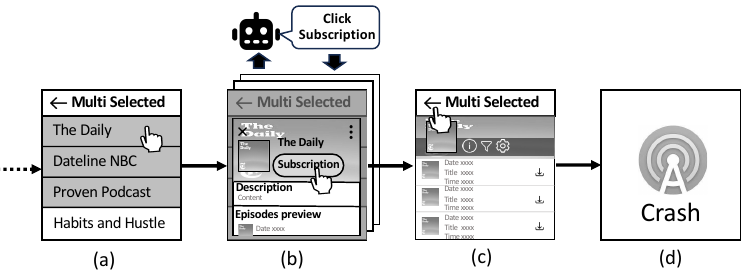}
    \caption{A real-world bug found by \hm in \textit{Antennapod}}
    \label{fig:case-study}
    \vspace{-0.5cm}
\end{figure}

We illustrate the benefits of our hybrid testing approach using a real-world bug\footnote{This bug is confirmed and fixed by the developers at \url{https://github.com/AntennaPod/AntennaPod/issues/7609}} from \textsf{AntennaPod}~\cite{antennapod2025}, a popular podcast manager app with 1M+ installations on Google Play. 
Figure~\ref{fig:case-study} shows the simplified bug-triggering path identified by our approach.
To manifest the bug, a user must first transit from a multi-selection state to a podcast preview page (Figure~\ref{fig:case-study}(a)), subscribe to the selected podcast (Figure~\ref{fig:case-study}(b)), and subsequently navigate back from the detail page (Figure~\ref{fig:case-study}(c)).
Upon returning, the app crashes unexpectedly (Figure~\ref{fig:case-study}(d)).
Notably, this bug is triggered only if the "Subscription" action is performed before navigating back. 
In contrast, returning without subscribing cancels the multi-select state, thus failing to trigger the bug.

\begin{figure*}[t]
    \centering
    \includegraphics[height=3cm,width=0.8\linewidth]{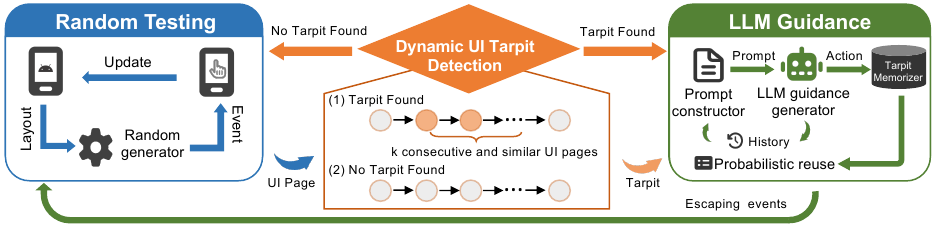}
    \caption{Workflow of Our Approach}
    \label{fig:overview}
    \vspace*{-0.4cm}
\end{figure*}

\noindent\textbf{Limitation of random testing.}
While random testing is efficient at fast and deep exploration, it remains semantics-oblivious and is prone to being trapped in UI tarpits.
As discussed earlier, UI tarpits are prevalent in random exploration.
Although human users can easily navigate to the next functional page, random testing often struggles and wastes resources exploring the current page.
For instance, page (b) in Figure~\ref{fig:case-study} represents a UI tarpit detected by our approach. 
The actual UI page contains 33 actionable UI widgets: 32 support 2 interaction types, while 1 supports 6 types, yielding a total action space of $S = 32 \times 2 + 1 \times 6 = 70$.
Among these 70 candidates, only 2 specific actions (\ie, clicking the "Subscription" or "Back" buttons) allow the app to transition away from the current page.
Crucially, only the "Subscription" action leads to new functional exploration, whereas the "Back" action merely returns to a previous page.
The remaining 68 actions result in the app trapping on the current page.
Consequently, the probability of a random event staying on page (b) is $p = \frac{68}{70} \approx 97.14\%$.
Following our definition of a UI tarpit ($k=8$ consecutive similar UI pages), the probability of random testing becoming trapped in this page is $p_t = p^{k} = (97.14\%)^8 \approx 79.30\%$, indicating being trapped in page (b) is a high-probability event. Indeed, we observe that random testing frequently encounters this tarpit and remains trapped there for steps far exceeding $k=8$, severely hindering testing efficiency.


In contrast, the probability $p_{b}$ of triggering the bug via pure random testing is extremely low.
To manifest the bug, random testing must generate a specific sequence: a "Subscription" event on page (b) followed by a "Back" event on page (c).
Given that page (c) contains 38 UI widgets (one supports 6 interaction types and others support 2 types), its action space size is $80=1*6+37*2$.
Thus, the combined probability is $p_{b} = \frac{1}{70} \times \frac{1}{80} \approx 0.18$\textperthousand.
Furthermore, the bug-triggering state is fragile: if the random testing selects "Back" on page (b) instead of "Subscription", it not only leaves the page but also cancels the multi-selection state, destroying the necessary precondition for the bug.
However, invoking an LLM on page (b) elevates the probability of clicking the "Subscription" button from $\frac{1}{70} \approx 1.43\%$ to nearly 100\%, as the LLM correctly identifies subscription as the core functionality and prioritizes this semantically relevant action.

\noindent\textbf{Our hybrid testing approach.} 
To this end, we leverage the semantic understanding of LLMs to guide exploration when random testing is trapped in a UI tarpit.  
In fact, the discovery of the bug in Figure~\ref{fig:case-study} was made possible by the joint contribution of random and LLM-guided exploration.
When random testing became trapped in page (b), the LLM intervened by suggesting the ``Subscribe'' action, which not only escaped the tarpit but also satisfied the precondition for the bug.
Subsequently, random testing resumed and clicked the ``back'' button, triggering the crash.
However, LLMs often bypass edge cases where many bugs are hidden. In this case, the LLM struggled to reach the multi-select state, which represents a boundary functional scenario in \textsf{AntennaPod} and thus is de-prioritized by the LLM.
This hybrid strategy combines the deep exploration capabilities of random testing with the semantic reasoning of LLMs, effectively exposing a broader range of latent bugs.

\section{LLM-Powered Random Testing}


\subsection{Overall Approach}
\label{sec:overall_approach}

Figure~\ref{fig:overview} illustrates the overall workflow of our approach.
Given an app under test (AUT), it initiates a random testing phase that continuously generates and executes random UI events.
Concurrently, a dynamic UI tarpit detector (\S\ref{sec:tarpit_detector}) monitors the sequence of executed UI pages to identify potential UI tarpits. 
Specifically, we classify a sequence of consecutive, visually similar UI pages as a tarpit.
Once a tarpit is detected, our approach switches to the LLM-guided exploration phase (\S\ref{sec:llm_guidance}).
In this phase, the system captures the current UI page and queries the tarpit memory to determine if the tarpit has been visited previously.
If it is a known tarpit, we probabilistically reuse prior successful escaping events.
Otherwise, we encode both the UI page(s) and the history of failed actions (if any) to construct a prompt that requests the LLM to suggest events capable of escaping the tarpit.
Upon successful escape, the approach resumes random testing.

Our approach interleaves these two complementary phases continuously throughout the testing process, as shown in Algorithm~\ref{alg:workflow}. In short, random testing drives the main testing loop, switching to LLM-guided exploration only when a UI tarpit is detected.

It takes the AUT ($\mathcal{A}$) as input and iterates to explore deep and diverse GUI states until the time budget is exhausted (Lines~3--15).
It first initializes both the tarpit memory $M$ and the GUI state sequence $S$, while capturing and appending the initial state $s$ of the app to $S$ (Line~2).
In the main testing loop, \textsf{hasTarpit} determines whether the app is trapped in a UI tarpit based on the last $k$ consecutive states (where $k$ represents the minimum window length required for detection, as detailed in \S\ref{sec:tarpit_detector}) (Line~4).
If no tarpit is detected, a random event is generated based on the current state $s$ and executed on $\mathcal{A}$ (Lines~5--6).
Subsequently, the new state $s$ is captured and appended to $S$ (Line~7).

If a tarpit is detected, it switches to LLM-guided exploration to generate an escaping event $e_{esc}$ within a maximum number of retries (Lines~9--10).
Specifically, \textsf{genEscapingEvent} integrates a probabilistic reuse mechanism with LLM-guided generation, employing an occlusion filtering algorithm to ensure the LLM accurately perceives the visible UI context.
After executing $e_{esc}$, it captures the new state $s$ and updates $S$ to verify if the tarpit has been successfully escaped.
If escaped, it reverts to random testing and records the effective escape event $e_{esc}$ along with its corresponding tarpit state (the preceding state in $S$) in memory $M$ (Lines~13--15), facilitating future escapes if the same tarpit is encountered again.


\begin{figure*}[t]
\removelatexerror  
    \begin{minipage}[t]{0.47\textwidth}
        \begin{algorithm}[H] 
            \footnotesize
            \DontPrintSemicolon
            \SetKwInOut{Input}{Input}
            \SetKwFunction{Main}{Main}
            \SetKwInOut{Const}{Const}
            \SetKwProg{Function}{Function}{:}{}
            \SetKwFunction{GetState}{getState}
            \SetKwFunction{RandGen}{genRandomEvent}
            \SetKwFunction{Execute}{execute}
            \SetKwFunction{IsTarpit}{hasTarpit}
            \SetKwFunction{ReuseGen}{genReuseEvent}
            \SetKwFunction{PromptGen}{genPrompt}
            \SetKwFunction{LLMGen}{genLLMEvent}
            \SetKwFunction{rand}{random}
            \SetKwFunction{filter}{getValidWidgets}
            \SetKwFunction{GetLeaves}{GetInteractiveLeaves}
            \SetKwFunction{Escape}{genEscapingEvent}
            \SetKwFunction{updateMemory}{updateTarpitMemory}
            \SetFuncSty{textsf}

            \Function{\Main($\mathcal{A}$)}{
                $M \gets \emptyset;$
                $s \gets$ \GetState($\mathcal{A}$); $S \gets [s]$\;
                
                \While{$not$ $timeout$}{
                    
                    \If{$\neg$\IsTarpit($S,k$)}{ 
                        $e \gets$ \RandGen($s$)\;
                        \Execute($e$)\;
                        $s \gets$ \GetState($\mathcal{A}$); $S \leftarrow S + [s]$\;
                    }\Else{
                        \tcp{Tarpit Found \& Escaping}
                        \For{ $q=0$ \KwTo \texttt{MAX\_RETRY}}{
                            $e_{esc} \gets$ \Escape($s,M$)\;
                            \Execute($e_{esc}$)\;
                            $s \gets$ \GetState($\mathcal{A}$); $S \leftarrow S + [s]$\;
                            
                            \If{$\neg$\IsTarpit($S,k$)}{
                                $M \gets M \cup \{ (S[|S|-2], e_{esc}) \}$\;
                    
                                \textbf{break}\;
                            }
                        }
                    }
                }
            }
 
            \caption{\small LLM-Powered Random Testing}
            \label{alg:workflow}
        \end{algorithm}
    \end{minipage}
    \hfill 
    \begin{minipage}[t]{0.47\textwidth}
        \begin{algorithm}[H] 
            \footnotesize 
            \DontPrintSemicolon 

            \SetKwInOut{Input}{Input}
            \SetKwInOut{Output}{Output}
            \SetKwInOut{Global}{Global}
            \SetKwInOut{Const}{Const}
            \SetKwProg{Function}{Function}{:}{}
            \SetKwFunction{calculateSimilarity}{CalculateSimilarity}
            \SetKwFunction{isUISimilar}{isUISimilar}
            \SetKwFunction{getScreenImage}{GetScreenImage} 
            \SetKwFunction{generatePerceptualHash}{PerceptualHash} 
            \SetKwFunction{calculateHammingDistance}{HammingDistance}
            \SetKwFunction{getHashLength}{len}
            \SetKwFunction{IsTarpit}{hasTarpit}
            \SetFuncSty{textsf}
    
            \Const{Sim threshold $\theta$}

            \Function{\IsTarpit($S,k$)}{
                $N \gets |S|$\;
                \If{$N < k$}{ \Return \textbf{False}\; }
                
                \tcp{Check last $k$ states}
               
                \For{$i \gets N-k$ \KwTo $N-2$}{
                    \If{$\neg$ \isUISimilar($S[i], S[i+1]$)}{
                        \Return \textbf{False}\;
                    }
                }
                \Return \textbf{True}\;
            }
            \BlankLine
            \BlankLine

            \Function{\isUISimilar($s, s'$)}{
                $hash \gets$ \generatePerceptualHash$(s)$\; 
                $hash' \gets$ \generatePerceptualHash$(s')$\; 
                $dist \gets$ \calculateHammingDistance($hash, hash'$)\;
                $score \gets 1.0 - (dist / $ \getHashLength($hash$))\;
                \textbf{return} $score \geq \theta$\;
            }

            \caption{\small UI Tarpit Detection}
            \label{algorithm:detect_tarpit}
        \end{algorithm}
    \end{minipage}
    \vspace*{-0.3cm}
\end{figure*}

\subsection{Dynamic UI Tarpit Detection}
\label{sec:tarpit_detector}

As discussed in \S\ref{sec:motivation}, UI tarpits often degrade testing efficiency. Intuitively, UI tarpit appears as a sequence of consecutive visually similar UI pages.
Based on this observation, we design an automatic detector to dynmically identify such tarpits during testing. 

\label{def:ui-tarpit}
A UI tarpit can be formally represented as a sequence of $k$ consecutive similar UI states $S=\left \langle s_0, s_1, \cdots, s_k \right \rangle $, where each transition $s_i\overset{e_i}{\rightarrow} s_{i+1}$ is triggered by a user event $e_i$, and $\forall 0\leq i<k, \text{ s.t. } Sim(s_i, s_{i+1})>\theta$. $Sim(\cdot)$ is similarity score, and $\theta$ is similarity threshold. 
The similarity score is calculated as the perceptual similarity~\cite{zhang2018unreasonable} between consecutive UI pages (\ie, UI states) using image hashing~\cite{dhash}, rather than the traditional view tree similarity (discussion in \S\ref{dis:image-based}).


The \emph{UI tarpit detector} continuously monitors the sequence of UI pages to determine whether the testing process has encountered or escaped a tarpit. As summarized in Algorithm~\ref{algorithm:detect_tarpit}, it takes a sequence of visited UI states $S=[s_0,s_1,\cdots]$ and a length threshold k as input. It iteratively examines the suffix of the sequence $S$ (\ie the last $k$ states) whether every adjacent pair is visually similar with \textsf{isUISimilar} function (Lines~5--8).
If any pair within the suffix are determined to be different, a \texttt{False} is immediately returned (Lines~6--7).
Otherwise, if all adjacent pairs among the last $k$ states are similar, \texttt{True} is returned, indicating the existence of a UI tarpit (Line~8).
A pair of UI pages is considered similar only if their similarity score exceeds the threshold $\theta$ (Lines~9--14),


\subsection{LLM-Powered Tarpit Escaping}
\label{sec:llm_guidance}
While efficient, random testing struggles to escape UI tarpits that require semantic interpretation.
Therefore, we design an LLM-guided exploration that leverages the LLM's capability to analyze state information and generate provide meaningful guidance actions to escape tarpits.
To balance the execution efficiency and exploration effectiveness, it probabilistically reuses the earlier escape action when encounters the same tarpit.
Formally, the escape policy determines the next event $e_{esc}$ based on the current state $s$ and the memory $M$:
\begin{equation}
\label{eq:dispatch}
\small
e_{esc} =
\begin{cases}
\text{Reuse}(M[s]), & \text{if } s \in M \land \zeta \le p \\
\text{LLMGuided}(s, H_{local}), & \text{otherwise}
\end{cases}
\end{equation}
\noindent where $\zeta \in [0,1]$ is a uniform random variable, $p$ is the reuse probability threshold, and $H_{local}$ denotes the local interaction history within the current escape phase.
This phase contains four components: \textit{Prompt Constructor}, \textit{LLM Guidance Generator}, \textit{Probabilistic Reuse Generator}, and \textit{Tarpit Memorizer}.

\subsubsection{Prompt Constructor}
\label{sec:prompt_constructor}

\begin{figure}[t]
    \centering
    \includegraphics[width=0.9\linewidth]{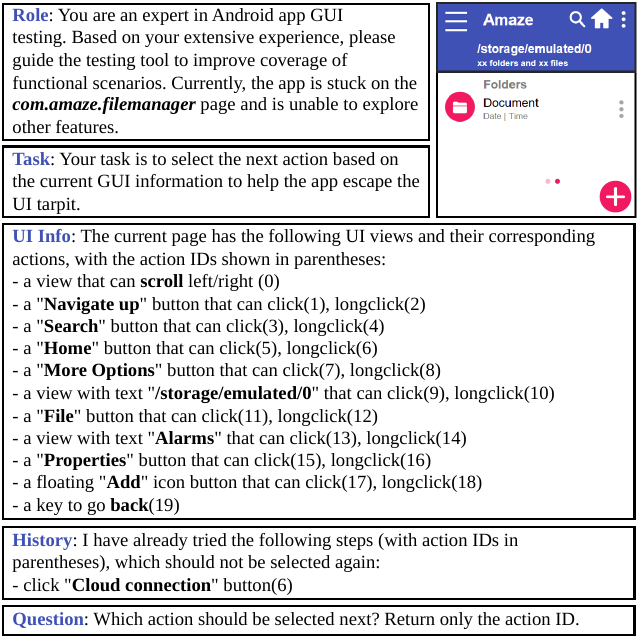}
    \caption{An illustration of the structured prompt construction. 
    }
    \label{fig:prompt}
    \vspace*{-0.6cm}
\end{figure}

It translates the GUI state into a structured textual representation that is comprehensible to the LLM.
Figure~\ref{fig:prompt} illustrates the structure of LLM prompt template, containing role, task, UI information, attempt history, and a question. 
The UI information is obtained through retrieving raw UI layout via the \textit{Android Accessibility Service}~\cite{accessbility}, and linearizing interactive widgets into a list of candidates. 
Specifically, we identify all enabled widgets and their supported interactions, augmenting descriptions with attributes such as \texttt{text}, \texttt{resource-id}, and \texttt{content-description}. 
Each widget is assigned a unique ID, enabling the LLM to signify its decision by returning a concise identifier instead of redundant text. To maintain a consistent spatial representation, we sort these widgets in a top-to-bottom and left-to-right order~\cite{liu2024make}. 

Sometimes, there are visual occlusion, including floating windows and pop-up layers, that prevents UI layout intepretation.
For example, a floating menu visually occludes the underlying file list, which may cause LLM-suggested action invalid (\eg, clicking a covered list item).
Therefore, we implement a spatial occlusion filter prior to prompt construction to make LLM focuses on truly visible context, as detailed in Algorithm~\ref{alg:widget}.
It first retrieves all interactive leaf nodes $W$ from the current state (Line~2), and iterates through each candidate widget $w$ to verify its spatial validity against other widgets (Lines~4--6).
It detects occlusion by examining whether the geometric center of $w$ falls within the bounding box of any other widget $w'$ (Line~7). Once overlapped, $w$ is flagged as covered and discarded (Lines 8-9) .
Ultimately, only the strictly visible widgets $W'$ are kept (Lines~10-12), effectively preventing the inclusion of misleading context that could cause invalid interactions.

To efficiently utilize the context window, we exclude the global execution trace and strictly limit the history to $H_{local}$, which comprises only the sequence of attempts made within the current tarpit instance, enabling the LLM to focus on relevant causal information.

\begin{algorithm}[t] 
    \scriptsize
    \DontPrintSemicolon
    \SetFuncSty{textrm}
    
    \SetKwInOut{Input}{Input}
    \SetKwProg{Function}{Function}{:}{}
    \SetKwFunction{filter}{getValidWidgets}
    \SetKwFunction{GetLeaves}{GetInteractiveLeaves}
    
    \Function{\filter{$S_t$}}{
         $W \gets$ \GetLeaves($S_t$)\;
         $W' \gets \emptyset$\;
         \For{$w \in W$}{
            $isCovered \gets \textbf{False}$\;
            \For{$w' \in W \setminus \{w\}$}{
                \If{$w.center \in w'.bounds$}{ \label{line:check_overlap}
                    $isCovered \gets \textbf{True}$\;
                    \textbf{break}\;
                }
            }
            \If{$\neg isCovered$}{
                $W' \gets W' \cup \{w\}$\;
            }
         }
         \Return $W'$\;
    }
    \caption{Widget Filtering}
    \label{alg:widget}
\end{algorithm}

\subsubsection{LLM Guidance Generator}
\label{sec:LLM_input}
Based on the constructed prompt with refined UI context, we design the LLM Guidance Generator to query the LLM and translate its high-level semantic decisions into executable system events.
To facilitate precise event execution, we maintain an \textsf{Action Space} that maps each interactive widget to its executable operations.
The \textsf{Action Space} $\mathbb{A}$ represents a discretized set of all executable GUI events on the current state, where each widget-level interaction is assigned a unique identifier to facilitate precise event execution.
Each event $e \in \mathbb{A} $ is a localized interaction, formally defined as a tuple $\langle id, bounds, type \rangle$, where $id$ is a unique identifier, $bounds$ denotes the coordinate area, and $type$ indicates the interaction type.
Upon receiving an LLM response (\eg, "Action ID: 6"), the generator performs a lookup to retrieve the corresponding metadata from the Action Space.

If the app persists in the tarpit after an escaping event, it invokes a feedback loop for strategy refinement.
In each iteration, the \textit{Prompt Constructor} updates the local history $H_{local}$ by appending the failed attempt, thereby instructing the LLM learn from the earlier mistakes and propose a new escaping event.
This interactive process continues until the tarpit is successfully escaped or the retry count reaches thelimit $q_{max}$ (default by 10). If all retries fail, the system terminates the LLM session and forces a "Back" operation to revert the app to the pre-tarpit state, resuming standard random exploration to prevent infinite stagnation.


\subsubsection{Tarpit Memorizer}
\label{sec:tarpit_memorizer}
It is a persistent registry for all encountered UI  tarpits.
Structurally, it maintains a collection of entries, where each entry is formalized as a tuple: \textit{<Tarpit ID, UI State, Action List>}, where \textit{Tarpit ID} is a unique identifier, the \textit{UI State} is the UI page of the tarpit, and the \textit{Action List} accumulates distinct actions that have successfully escaped this specific tarpit.
Upon detection of a tarpit, the memorizer queries the registry to determine whether the current UI state has been previously recorded.
This retrieval process employs the perceptual hashing metric defined in Algorithm~\ref{algorithm:detect_tarpit}, but enforces a significantly stricter similarity threshold $\theta_{mem}=0.99$.
This high threshold is deliberately chosen to ensure that historical actions are only reused on virtually identical states, guaranteeing safety and applicability, while accommodating minimal rendering variations (\eg, system clock or battery icon changes).
If a record detected(\ie, a ``visited'' tarpit), the associated history is forwarded to the probabilistic reuse generator to facilitate immediate escape.
Otherwise (\ie, a "new" tarpit), the system proceeds to the LLM-guided mode; once the tarpit is successfully escaped, the memorizer will store this effective solution.

\subsubsection{Probabilistic Reuse}
\label{sec:reuse_input}
While LLM guidance effectively mitigates UI tarpits, it takes high computational overhead.
To balance testing efficiency with state exploration, the probabilistic reuse generator employs a stochastic dispatch strategy.
Upon detecting a previously encountered UI tarpit, the generator activates the reuse mode with a high probability ($p=0.8$), randomly sampling a candidate from the set of successful actions recorded for the current tarpit instance. Otherwise, the LLM Guidance Generator is invoked, in an aim to generate potentially new and diverse escaping events.
We ground this threshold in the classic \textit{exploration-exploitation trade-off}: a dominant probability is assigned to exploitation to maximize the utility of cost-intensive LLM guidance, while the remaining ($1-p=0.2$) is reserved for exploration to foster the discovery of alternative escape paths.
This strategy effectively reduces redundant API costs while preserving the diversity of behavior.

\section{Implementation}



\label{sec:implement}
We realize our idea on two different random testing tools and obtain \hm and \hd. Both of them utilize GPT-4o as their underlying LLM.
\hm is built upon \rmk, an enhanced version of the official Android \mk~\cite{monkey}.
\mk is widget-oblivious, injecting events at random coordinates without considering the UI structure.
To alleviate this, we leverage \ui~\cite{uiautomation} to obtain widgets and their supported events.
In practice, such widget-aware exploration has been shown to statistically improve code coverage~\cite{zeng2016automated}.
\hd is based on \rd, an implementation of a random exploration policy within \db~\cite{li2017droidbot}, a widely used academic AIG tool.
\hd utilizes \utwo~\cite{uiautomator2} for UI manipulation.

We adopted the OpenCV library~\cite{opencv} for image similarity calculation.
The similarity threshold $\theta$ is determined empirically through a pilot study conducted on a subset of 4 apps, where we evaluated values in the range $[0.90, 0.99]$ and manually inspected the detected tarpits.
We observed that lower values (\eg, $\theta = 0.91$) led to over-approximation by merging distinct UI pages and higher values (\eg, $\theta = 0.98$) were overly sensitive, failing to group perceptually identical pages.
We therefore set $\theta = 0.95$ to achieve a balance between trade-off for maintaining detection accuracy.
The sequence length threshold $k$ is set to 8.

\section{Evaluation}
\label{sec:evaluation}

\label{sec:research_questions}
Our evaluation aims to answer four research questions:

\begin{itemize}[leftmargin=*]
\item \textbf{RQ1 (Code Coverage):} Compared to existing automated testing techniques, how effective is our approach in code coverage?
    
\item \textbf{RQ2 (Bug Detection):} Compared to existing automated testing techniques, how effective is our approach in detecting bugs?

\item \textbf{RQ3 (Escape Effectiveness):} How effective is our approach in detecting and escaping UI tarpits? And how much does a successful escape improve code coverage?

\item \textbf{RQ4 (Ablation Study):} How do individual components contribute to the overall effectiveness of our hybrid strategy?

\end{itemize}



\subsection{Evaluation Setup and Method}
\label{sec:evaluation_setup}

\noindent\textbf{App Subjects.} We create a benchmark of 13 apps, containing (1) eight representative, open-source apps from prior work in automated Android GUI  testing~\cite{su2021benchmarking,xiong2024general,mao2016sapienz,su2021fully}; (2) four sourced from Google Play to enhance functional diversity; and (3) \textit{WeChat}~\cite{wechat2025}, a large-scale commercial app characterized by a highly complex UI.
We use open-source apps as our primary subjects for collecting precise code coverage data and submiting bug reports.
Table~\ref{tab:apps} summarizes the statistics of these apps.
To ensure generalizability, we also evaluate on two additional benchmarks in \S~\ref{extend_eval}.

\noindent\textbf{Baseline Selection.}
\label{baseline}
To evaluate from multiple perspectives, We use 7 baselines from three groups: traditional AIG tools (Group A), tarpit-specialized tools (Group B), and LLM-based testing tools
(Group C).

\begin{itemize}[leftmargin=*,topsep=0pt]
    \item \textit{\textbf{Baseline Group A.}}
    We include two widely-used industrial tools: \mk~\cite{monkey}, the Android random testing tool; and \fb~\cite{lv2022fastbot2}, a state-of-the-practice reinforcement learning-based tool from ByteDance. 
    To ensure a fair comparison and isolate the impact of our strategy, we also introduce two variants (see \S\ref{sec:implement}):\rmk, a widget-based adaptation of Monkey; and \rd, an extended version of \db configured with a random strategy to serve as widget-level random baselines.
    
    \item \textit{\textbf{Baseline Group B.}}
    We compare against \au~\cite{khan2024aurora}, a state-of-the-art tool designed to escape UI tarpits via heuristic rules.
    While \vet~\cite{wang2021vet} is the first work to define the term \textit{UI exploration tarpits}, we omit it as it aims to \textit{prevent} rather than \textit{escape} tarpits, and \au has demonstrated superior performance over Vet.

    \item \textit{\textbf{Baseline Group C.}}
    We select two representative LLM-based tools: \gptd~\cite{liu2024make}, which employs a step-by-step LLM decision-making strategy; and \llmd~\cite{wang2025llmdroid}, a most recent work that integrates LLMs into existing AIG tools. 
    As \gptd does not offer an official replication package, we faithfully reconstructed it with all modules from its open-access repository~\cite{gptdroid2023}.
\end{itemize}

\begin{table}[t]
\centering
\caption{App subjects used in our experiment (K=1,000, M=1,000,000).}
\begin{adjustbox}{max width=0.95\linewidth}
\begin{tabular}{l l r r r r>{\centering\arraybackslash}p{2.5cm}}
\toprule
\textbf{App Name}  & \textbf{App Feature}  & \textbf{Stars} & \textbf{Downloads} & \textbf{LOC} & \textbf{APK Size} \\ \hline
Amaze              & File Manager              & 5.3K           & 1M+                & 111,214      & 11.53MB       \\
AnkiDroid          & Flashcard Learning          & 8.6K           & 10M+               & 470,803      & 105.91MB      \\
AntennaPod         & Podcast Manager     & 6.4K           & 1M+                & 641,889      & 11.53MB       \\
Chess              & Casual             & 468            & 500K+              & 59,230       & 7.65MB        \\
Feeder             & RSS Reader       & 1.6K           & 100K+              & 152,340      & 60.82MB       \\
MyExpenses         & Expense Tracking            & 820            & 1M+                & 223,082      & 45.09MB       \\
NewPipe            & Video Manager     & 31.4K          & 7.6M+              & 317,897      & 11.53MB       \\
Omni-Notes         & Note Manager       & 2.7K           & 100K+              & 73,100       & 7.65MB        \\
OwnTracks          & Location Tracking    & 1.4K           & 100K+              & 122,323      & 13.63MB       \\
RedReader          & Social Discussion         & 2K             & 100K+              & 171,039      & 9.02MB        \\
SimpleAlarm & Time Manager              & 510            & 1M+                & 92,145       & 7.97MB        \\
Wikipedia          & Knowledge Reference & 2.4K           & 50M+               & 557,744      & 77.59MB       \\ 
WeChat             & Messaging \& Social      & -              & 100M+              & -            & 243.52MB       \\
\bottomrule
\end{tabular}
\end{adjustbox}
\label{tab:apps}
\vspace*{-0.3cm}
\end{table}

\noindent\textbf{Environmental Configuration.}
To mitigate randomness, we executed each tool five times on all apps. Following prior studies~\cite{su2017guided,wang2018empirical}, we set a time budget of 3 hours per run to ensure sufficient exploration depth.
All experiments were conducted on a 64-bit Ubuntu 22.04 machine (128-core AMD EPYC 7742 CPU, 256 GB RAM) using the official Android emulator configured as a Google Pixel 4 device (Android 11, 4-core CPU, 4 GB RAM) except for \textit{WeChat}. The experiments on \textit{WeChat} were conducted on a real device (SHARKKLE-A0, Android 11).
All baselines followed their original configurations, except that \gptd and \llmd were standardized to use GPT-4o for a fair comparison.


\noindent\textbf{Evaluation method of RQ1.}
We collected coverage at four levels: \textit{Line, Branch, Method,} and \textit{Class} via $\mathtt{JaCoCo}$~\cite{jacoco}, a widely used instrumentation tool. We also measure activity coverage by calculating the ratio of visited activities to the total set defined in the $\mathtt{AndroidManifest.xml}$ file. 
For \textit{WeChat}, where source code is unavailable, we use activity coverage as a proxy metric.

\noindent\textbf{Evaluation method of RQ2.}
We recorded unique crashes triggered during testing, de-duplicated by stack traces from $\mathtt{Logcat}$~\cite{logcat,su2017guided} logs.
In addition, we run \hd, \hm, \mk, \fb, and \db 10 times for statistical comparisons. 
We employed the Wilcoxon rank-sum test~\cite{wilcoxon1992individual} and Vargha and Delaney's $\hat{A}_{12}$~\cite{vargha2000critique} to evaluate statistical significance (significant when $p\text{-val}<0.05$) and effect size (effective when $\hat{A}_{12}>0.5$), respectively.
Moreover, we report the uncover new crashes of \hm across 10 independent runs.
\begin{table}[t]
    \centering
    \setlength{\tabcolsep}{1.2pt} 
    \scriptsize 
    
    \captionsetup[subtable]{font=footnotesize, labelformat=simple}
    \renewcommand\thesubtable{(\alph{subtable})}
    \caption{Coverage comparison among different baselines.}
    \label{tab:overall_coverage}

    \begin{subtable}{0.38\linewidth}
        \centering
        \caption{Baselines A}
        \label{tab:rq1_cov}
        \vspace{-0.2cm}
        \begin{tabularx}{\linewidth}{lYYYYYY}
        \toprule
        Cov(\%) & D* & F & M & M* & Hd & Hm \\ \midrule
        Line & 32.1 & 34.0 & 29.9 & 36.5 & \textbf{40.3} & \textbf{43.1} \\
        Branch & 21.1 & 23.5 & 19.8 & 24.8 & \textbf{28.0} & \textbf{30.1} \\
        Method & 34.9 & 36.7 & 32.8 & 39.8 & \textbf{43.7} & \textbf{46.8} \\
        Class & 43.6 & 45.8 & 41.9 & 49.0 & \textbf{52.4} & \textbf{55.2} \\
        Activity & 37.3 & 37.6 & 36.3 & 41.5 & \textbf{43.2} & \textbf{45.7} \\ \bottomrule
        \end{tabularx}
    \end{subtable}
    \hfill
    \begin{subtable}{0.29\linewidth}
        \centering
        \caption{Baselines B}
        \label{tab:rq2_cov}
        \vspace{-0.2cm}
        \begin{tabularx}{\linewidth}{lYYY}
        \toprule
        Cov(\%) & Hm & Hd & A \\ \midrule
        Line & \textbf{43.1} & 40.3 & 24.1 \\
        Branch & \textbf{30.1} & 28.0 & 15.8 \\
        Method & \textbf{46.8} & 43.7 & 25.6 \\
        Class & \textbf{55.2} & 52.4 & 35.2 \\
        Activity & \textbf{45.7} & 43.2 & 27.8 \\ \bottomrule
        \end{tabularx}
    \end{subtable}
    \hfill
    \begin{subtable}{0.29\linewidth}
        \centering
        \caption{Baselines C}
        \label{tab:rq3_cov}
        \vspace{-0.2cm}
        \begin{tabularx}{\linewidth}{lYYYY}
        \toprule
        Cov(\%) & Hm & Hd & LLM & GPT \\ \midrule
        Line & \textbf{43.1} & 40.3 & 33.9 & 16.9 \\
        Branch & \textbf{30.1} & 28.0 & 21.6 & 10.6 \\
        Method & \textbf{46.8} & 43.7 & 35.7 & 19.4 \\
        Class & \textbf{55.2} & 52.4 & 46.3 & 27.4 \\
        Activity & \textbf{45.7} & 43.2 & 31.4 & 21.4 \\ \bottomrule
        \end{tabularx}   
    \end{subtable}
    \vspace{-0.1cm}
\end{table}

\begin{figure}[t]
    \centering
    \includegraphics[height=4.5cm]{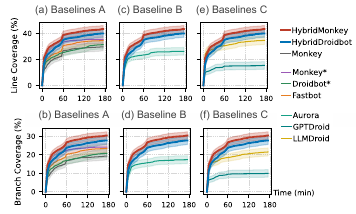}
    \caption{Average line and branch coverage growth on 12 apps across three baseline groups (3-hour testing). 
    }
    \vspace{-0.5cm}
    \label{fig:coverage-avg}
\end{figure}

\begin{figure*}[t]
    \centering
    \includegraphics[width=0.9\linewidth]{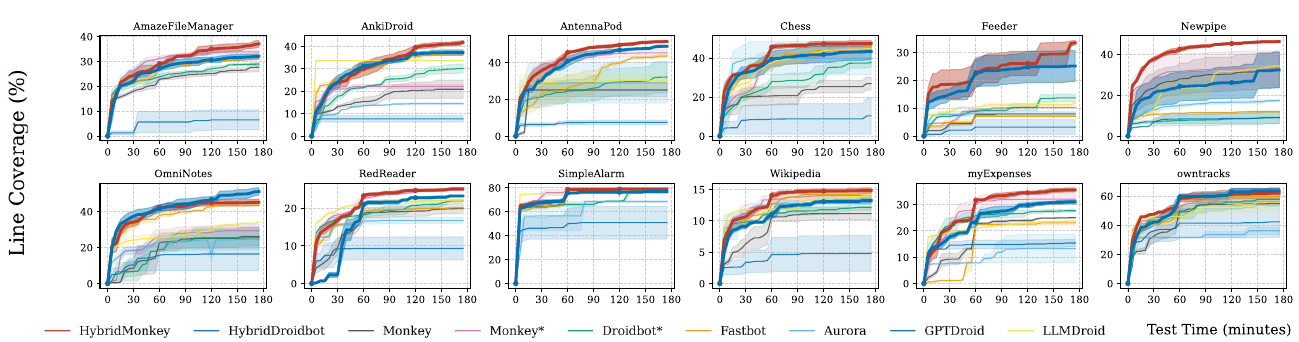}
    \caption{Line coverage over time of 9 tools across 12 apps.
    }
    \label{fig:coverage-apps}
    \vspace*{-0.4cm}
\end{figure*}

\begin{figure*}[!t]
    \centering
    \includegraphics[width=0.9\linewidth]{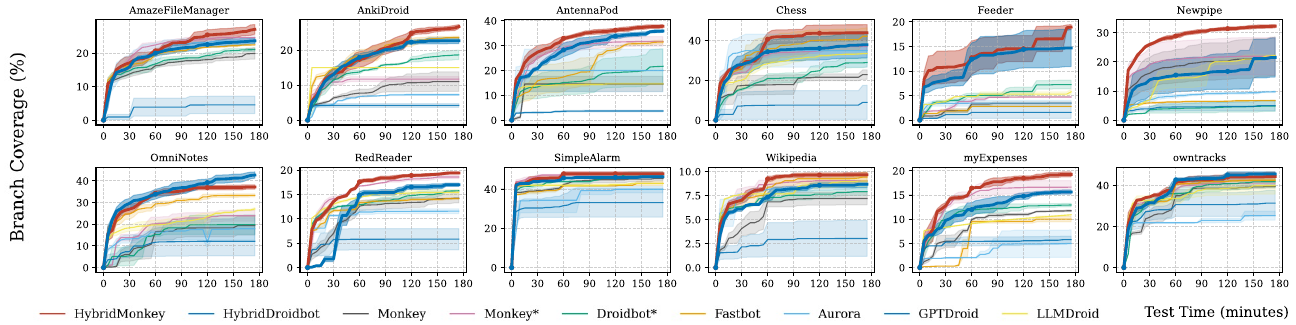}
    \caption{Branch coverage over time of 9 tools across 12 apps.
    }
    \label{fig:branch-coverage-apps}
    \vspace*{-0.4cm}
\end{figure*}

\noindent\textbf{Evaluation method of RQ3.}
As we lack ground-truth of UI tarpits, we use following metrics.


\noindent\textit{1) Tarpit Detection Precision (TDP).}
Due to the large volume of interaction logs, we randomly sampled one execution trace per app (out of five runs) and manually verified all reported tarpits.
A detection is considered a True Positive if the UI layout and functional state remain substantially unchanged, as well as without new actionable elements for the preceding $k=8$ consecutive steps. This verification window aligns with the similarity threshold used in our algorithm.

\noindent\textit{2) Escape Success Rate (ESR).} 
Using the same sampled traces, we manually verified whether the LLM-generated actions successfully escaped the tarpits.
We report a success for (a) \textit{Valid Return} (\ie returning to the parent node when exploration is exhausted) and (b) \textit{New Path Discovery} (\ie triggering a significant change in page state or interactive elements).

\noindent\textit{3) First-Attempt Escape Rate (FAER).}
We calculate it on 5 runs across 12 apps, by counting escapes achieved solely by the initial LLM query.
A success is reported when there is visual discrepancies between pre- and post-action screenshots, which indicate a valid state transition.

\noindent\textit{4) Post-Escape Coverage Contribution (PEC).}
To quantify the impact of escapes on coverage, we analyze the temporal correlation between LLM queries and ``coverage inflection points'' (\ie, instances of new line coverage).
Specifically, we calculate the numbers of LLM queries within the 50-second window (\ie, five 10-second intervals) immediately preceding each inflection point. We report the average results across five runs. 


\noindent\textbf{Evaluation method of RQ4.}
RQ4 aims to assess the contribution of each component. We conduct an ablation study comparing the full approach against two variants: (1) \textit{w/o Reuse} (Probabilistic Reuse disabled), and (2) \textit{w/o LLM} (LLM guidance disabled).
We evaluate these configurations on both \hm and \hd in terms of code coverage.

\vspace{-5pt}
\subsection{RQ1: Code Coverage}
\label{sec:rq1_coverage}

\noindent\textbf{Baseline Group A.}
Table~\ref{tab:rq1_cov} and Figure~\ref{fig:coverage-apps}\&\ref{fig:branch-coverage-apps} present the average activity and code coverage results.
Our tools consistently outperform all baselines across all metrics and apps.
On average, \hd and \hm achieve 40.3\%--43.1\% line coverage, surpassing baselines (29.9\%--36.5\%).
In particular, \hm exceeds \mk by 44.1\% and 51.8\% in line and branch coverage, as \mk's coordinate-based generation often yields invalid events.
Compared to widget-based \rmk, \hm maintains a 10.1\%--21.4\% lead across all coverage metrics.
Similarly, \hd outperforms \rd by 15.8\%--32.4\%.
These results indicate that our tool effectively broadens the exploration scope, thereby increasing the potential for bug discovery.
We provide a detailed analysis in \S~\ref{sec:rq3} to verify if this improvement stems from the tarpit escape mechanism.

Figure~\ref{fig:coverage-avg} (a)\&(b) illustrate the line/branch coverage growth for baseline group A. We can see that our tools outperformed \fb, \mk and \rd from around $10^{\mathrm{th}} $ minute onwards, finally achieving the highest overall coverage at the end of the testing budget. 
Notably, during the first 60 minutes, \rmk exhibited competitive performance, closely trailing our \hd and outperforming the other three baselines, yet it consistently remained inferior to \hm.
Our tools has narrower regions (standard deviation), indicating better stability. 

\label{aurora-tarpit}
\noindent\textbf{Baseline Group B.} 
Table~\ref{tab:rq2_cov} shows that \au achieves significantly lower coverage than our approach.
Specifically, \hm's line coverage (43.1\%) is nearly double that of \au (24.1\%).
The growth curves in Figure~\ref{fig:coverage-avg} (c)\&(d) further illustrate that \au's coverage tends to plateau early.
The reason is that \au relies on eight fixed pattern matching, lacking generalization to diverse UI designs (\eg, our motivating example in \S\ref{sec:motivation}).
Unlike \au, which resets exploration after three failed heuristic attempts, our approach dynamically adapts to unseen UI states via LLM-driven semantic reasoning.

\noindent\textbf{Baseline Group C.}
Table~\ref{tab:rq3_cov} shows that \hm achieves the highest coverage, outperforming \llmd by 27.3\% (line) and 39.5\% (branch). \gptd lags significantly (16.9\% line coverage).
This gap may stem from the heavy reliance on LLMs in these baselines.
First, \gptd uses LLMs for every step, while \llmd depends on them for multiple stages, which creates a long dependency chain that is vulnerable to incomplete UI metadata (\eg, missing text or resource IDs).
Consequently, perception inaccuracies can propagate downstream and reduce testing effectiveness.
Second, these limitations are likely amplified during our 3-hour experiments compared to their original 1-hour experiments.
The cumulative latency and context limits may hinder their performance over time.
Conversely, our approach invokes LLMs only for tarpits, ensuring sustained growth and stability.
Figure~\ref{fig:coverage-avg}~(e)\&(f) show that our approach maintain a sustained upward trend, whereas \llmd and \gptd plateau early, typically within 60 mins.

\noindent\textbf{Results on WeChat.}
\textit{WeChat} is a massive industrial app with 1,988 activities.
Given that previous results established our superiority over other baselines, here we focus on the enhancement to the random strategy.
We compared \hm directly against \rmk to verify this improvement in an industrial setting. 
\hd was excluded due to infrastructure constraints within WeChat's internal environment; unlike the host-dependent \hd, \hm enables direct, on-device execution.
\hm covered 122 activities on average, surpassing \rmk's 116.
This confirms our hybrid approach effectively improves exploration in complex, real-world scenarios.

\begin{figure}[t]
    \centering
    \includegraphics[width=0.8\linewidth]{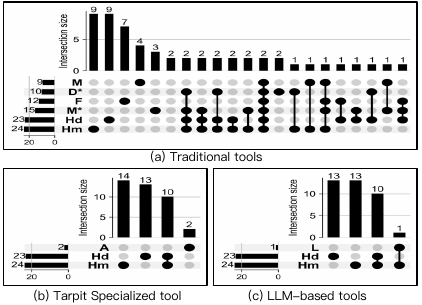}
    \caption{Intersection analysis of crashes across different baseline tools. \textmd{\footnotesize{(Horizontal bars indicate total crashes per tool; vertical bars show the number of crashes in each intersection.)}}}
    \label{fig:upset}
    \vspace{-0.6cm}
\end{figure}

\vspace{-5pt}
\subsection{RQ2: Bug Detection}
\label{sec:rq2}
\noindent\textbf{Baseline Group A.} 
Figure~\ref{fig:upset}~(a) presents the crash analysis via an UpSet plot. The horizontal bars on the left show the total number of unique crashes detected by each tool. \hm and \hd detect the most unique crashes (24 and 23), significantly surpassing \rmk (15), \fb (12), and \mk (9). \label{lesson1}
Beyond total counts, intersection analysis (vertical bars) reveals whether our approach reveals bugs missed by other techniques. As shown in the first two columns, \hm and \hd each detect 9 exclusive crashes, whereas \fb and \rmk find only 7 and 3, respectively.
This enhanced detection capability stems from our hybrid strategy's superior exploration efficiency.
By ensuring both exploration breadth and depth, our approach finds unique bugs that remain inaccessible to the baselines.

\noindent\textbf{Baseline Group B.}
As shown in Figure~\ref{fig:upset} (b), \hm and \hd detected 24 and 23 unique crashes, respectively, while Aurora identified only 2 crashes in total. In terms of exclusive detections, \hm and \hd successfully identified 14 and 13 unique crashes that were missed by Aurora. 

\noindent\textbf{Baseline Group C.}
Figure~\ref{fig:upset} (c) shows our tools (24 and 23 crashes) significantly outperforming \llmd (1 crash), with \gptd detecting none.
We attribute the limited effectiveness of \llmd to two main factors.
First, \llmd prioritizes rapid coverage expansion over deep exploration.
When coverage stagnates, it immediately shifts its focus to a different state subspace.
However, critical crashes are often located deeply behind these stagnation points.
Second, \llmd utilizes LLM guidance to target uncovered functions sequentially.
It identifies major functional units and executes them one by one based on priority.
This isolated execution interrupts the continuous interaction chain.
Consequently, it fails to accumulate the complex states required to trigger deep crashes.

\begin{table}[t]
    \centering
    \footnotesize 
    
    \begin{minipage}[t]{0.68\linewidth} 
        \centering
        \caption{Statistical comparison.}
        \label{tab:stat-bug}
        
        \setlength{\tabcolsep}{1.5pt} 
        \begin{tabular}{|l|cc|cc|cc|}
        \hline
        \multirow{2}{*}{\textbf{App}} & \multicolumn{2}{c|}{\textbf{Hm vs M*}} & \multicolumn{2}{c|}{\textbf{Hm vs F}} & \multicolumn{2}{c|}{\textbf{Hd vs D*}} \\ \cline{2-7} 
                                      & p-val & $\hat{A}_{12}$ & p-val & $\hat{A}_{12}$ & p-val & $\hat{A}_{12}$ \\ \hline
        SimpleAlarm & N/A & N/A & N/A & N/A & N/A & N/A \\
        AmazeFile   & 0.52 & 0.64$\uparrow$ & \textbf{0.01*} & \textbf{1.00$\uparrow$} & \textbf{0.01*} & \textbf{1.00$\uparrow$} \\
        AnkiDroid   & N/A & N/A & N/A & N/A & N/A & N/A \\
        AntennaPod  & \textbf{0.02*} & \textbf{0.96$\uparrow$} & \textbf{0.03*} & \textbf{0.90$\uparrow$} & \textbf{0.01*} & \textbf{1.00$\uparrow$} \\
        Chess       & 0.06 & 0.86$\uparrow$ & 1.00 & 0.50 & \textbf{0.03*} & \textbf{0.92$\uparrow$} \\
        feeder      & \textbf{0.01*} & \textbf{1.00$\uparrow$} & 0.07 & 0.80$\uparrow$ & 0.42 & 0.60$\uparrow$ \\
        MyExpenses  & 0.18 & 0.70$\uparrow$ & N/A & N/A & 0.18 & 0.70$\uparrow$ \\
        Newpipe     & \textbf{0.02*} & \textbf{0.96$\uparrow$} & \textbf{0.01*} & \textbf{1.00$\uparrow$} & \textbf{0.01*} & \textbf{1.00$\uparrow$} \\
        OmniNotes   & \textbf{0.02*} & \textbf{0.92$\uparrow$} & \textbf{0.01*} & \textbf{1.00$\uparrow$} & \textbf{0.02*} & \textbf{0.92$\uparrow$} \\
        Owntrack    & \textbf{0.01*} & \textbf{1.00$\uparrow$} & \textbf{0.03*} & \textbf{0.92$\uparrow$} & 1.00 & 0.52$\uparrow$ \\
        RedReader   & 0.42 & 0.60$\uparrow$ & N/A & N/A & 0.42 & 0.60$\uparrow$ \\
        WikiPedia   & 0.12 & 0.76$\uparrow$ & 0.42 & 0.60$\uparrow$ & 0.12 & 0.76$\uparrow$ \\ \hline
        \end{tabular}
        \vspace{2pt}
        \begin{flushleft}
            \hspace{0pt} 
            \scriptsize \textit{Note:} $p$-val $<$ 0.05 (* bold); $\hat{A}_{12} > 0.5$ ($\uparrow$) favors us. N/A: no crash.
        \end{flushleft}
    \end{minipage}%
    \hfill 
    \begin{minipage}[t]{0.29\linewidth} 
        \centering
        \caption{FAER.}
        \label{tab:faer}
        
        \setlength{\tabcolsep}{3pt}
        \begin{tabular}{|l|c|}
            \hline
            \textbf{Subject} & \textbf{FAER} \\ \hline
            SimpleAlarm & 93.4\% \\
            AmazeFile   & 83.4\% \\
            AnkiDroid   & 73.4\% \\
            AntennaPod  & 84.9\% \\
            Chess       & 90.0\% \\
            Feeder      & 68.4\% \\
            MyExpenses  & 89.1\% \\
            NewPipe     & 83.7\% \\
            OmniNotes   & 67.2\% \\
            OwnTracks   & 65.8\% \\
            RedReader   & 90.7\% \\
            WikiPedia   & 81.9\% \\ \hline
            \textbf{Overall} & \textbf{82.6\%} \\ \hline
        \end{tabular}
    \end{minipage}
    \vspace{-0.5cm}
\end{table}

\noindent\textbf{Statistical significance and effect size.}
Table~\ref{tab:stat-bug} reports the $p$-values and $\hat{A}_{12}$ effect sizes to verify our improvements.
We focus our statistical analysis on \rmk, \fb, and \db, as they are the only baselines detecting over 10 cumulative crashes, ensuring sufficient data for meaningful comparison.
\textit{SimpleAlarm} and \textit{AnkiDroid} are excluded due to zero crashes across all tools.
For the remaining subjects, \hm significantly outperforms \rmk and \fb on 5 subjects each, while \hd shows significant improvements over \db on 5 subjects.
The $\hat{A}_{12}$ values consistently exceed 0.5, with even non-significant cases often exhibiting large effect sizes (above 0.80). This occasional lack of strict significance likely results from the inherent variance of random testing and bug scarcity in stable apps, which reduces statistical power. Nonetheless, the consistent effect sizes confirm that our strategy effectively improves bug detection.

\noindent\textbf{Practical utility on real-world apps.}
Of the \totalBug unique bugs found by our approach, \reportBug previously unknown issues were submitted to developers.
Among them, \fixAndConfirm have been fixed or confirmed (\fixBug fixed and \confirmBug confirmed), while the rest remain under review. 
On \textit{WeChat}, \hm identified 5 unique crashes (vs. 3 by \rmk), demonstrating its robustness in detecting failures within complex industrial applications.

\subsection{RQ3: Escape Effectiveness}
\label{sec:rq3}

\noindent\textbf{Tarpit Detection Precision (TDP).} 
Our detection mechanism achieved 97.41\% precision (582/598).
We further analyzed the 16 false positives and identified two primary causes: 
\begin{itemize}[leftmargin=*]
    \item[1)] \textit{Low Contrast in Dark Mode}, where the algorithm showed reduced sensitivity to subtle visual changes in dark-themed interfaces;
    \item[2)] \textit{Transient Loading States}, where screenshots were captured during content loading (e.g., blank or spinner pages), leading to incorrect similarity judgments due to lack of visual information.
\end{itemize}

\noindent\textbf{Escape Success Rate (ESR).} 
On the 582 validated tarpits, our strategy achieved a 72.85\% (424/582) success rate. 
To understand the limitations, we analyzed the 158 failed instances and categorized the root causes into three primary types:

\begin{itemize}[leftmargin=*]
    \item \textbf{Incomplete Accessibility Information (64.56\%, 102/158):} \label{lesson4}
    The majority of failures stem from insufficient UI semantic information derived from the Layout tree. 
    First, the current layout analysis captures structural data but misses image-level details. 
    Second, many Android components contain empty attributes, hindering the LLM's understanding of the interface. 
    Consequently, this information gap leads to ineffective action generation, such as the LLM suggesting a ``long-press'' on a text view that actually requires a ``click''.
    
    \item \textbf{Random Strategy Interference (18.99\%, 30/158):} 
    In these cases, the LLM successfully initiated an escape action (e.g., clicking ``More Options'' to open a menu), but the subsequent random exploration failed to interact with the newly revealed controls. This discontinuity prevented the tool from completing the escape sequence, causing the app to revert to the tarpit state.
    
    \item \textbf{Text Input Constraints (16.46\%, 26/158):} 
    Failures also occurred when escaping required specific data inputs, particularly in Login or Registration scenarios. While our approach handles single text fields effectively, it struggles with complex multi-field dependencies that demand context-aware structured input.
\end{itemize}

Notably, the high proportion of missing accessibility information ($\approx$~65\%) indicates that the primary bottleneck lies in UI metadata availability rather than LLM reasoning (as discussed in \S~\ref{dis:lesson3}).


\noindent\textbf{First-Attempt Escape Rate (FAER).}
Table~\ref{tab:faer} shows our approach achieves an overall FAER of 82.55\%.
High rates in apps like \textit{Alarm} (93.38\%) and \textit{RedReader} (90.71\%) indicate that for most tarpit scenarios, the LLM is capable of identifying the critical exit interaction (\eg, "Back" or "Cancel") in a single inference step. This efficiency is crucial for large-scale testing, as it resolves most stagnation issues instantly and minimizes time spent in non-productive states.

\begin{figure}[t]
    \centering
    \includegraphics[width=\linewidth]{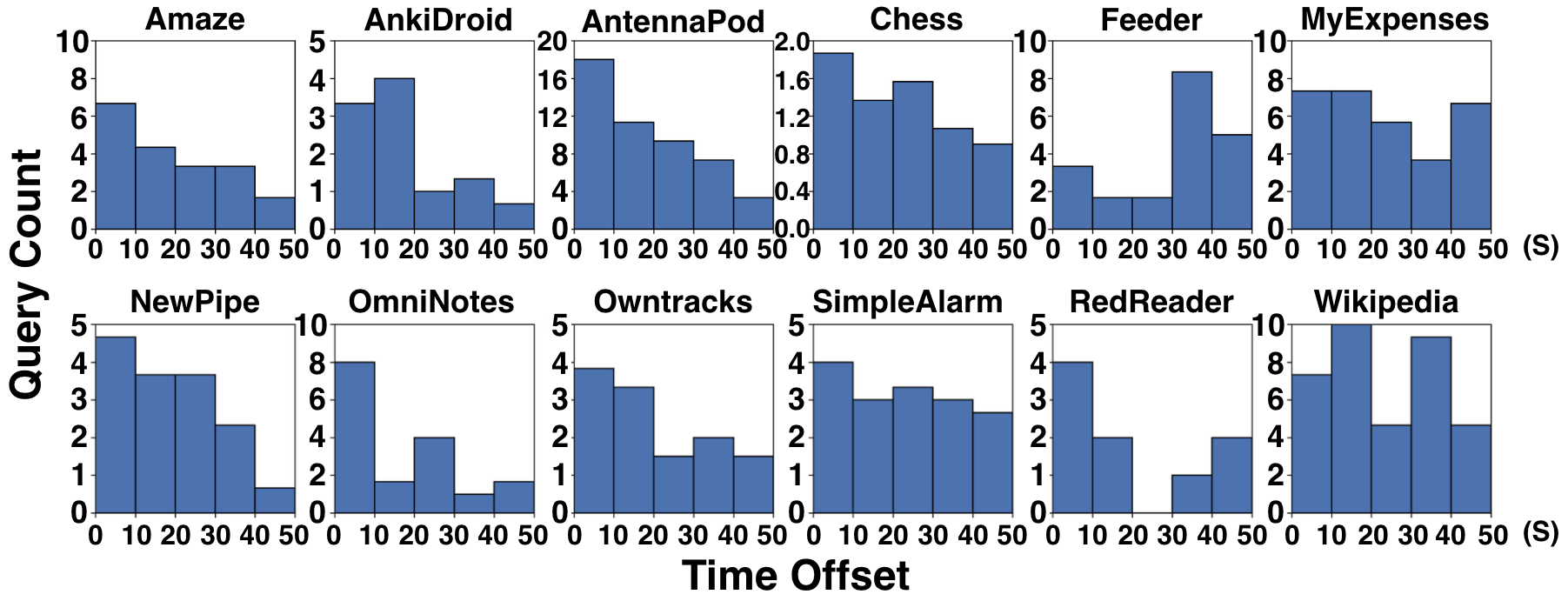}
    \caption{LLM query distribution before coverage gain points}
    \label{fig:llm-query-distribution}
    \vspace{-0.4cm}
\end{figure}

\noindent\textbf{Post-Escape Coverage Contribution (PEC).}
Figure~\ref{fig:llm-query-distribution} shows the average number of LLM queries before the line coverage increment of a test run, where the x-axis denotes the time interval before the increment. 
Most increments occur within 20s of a query, confirming that LLM-guided escaping effectively drives new exploration.
A notable exception is \textit{Feeder}, where coverage gains tend to occur in the 30-40s window after LLM queries. This delay is likely due to asynchronous feed loading and deferred UI rendering, where coverage improvement requires additional random events or background task completion. Nonetheless, the temporal correlation supports the causal role of LLM guidance in driving exploration forward.

\begin{table}[t]
    \centering
    \footnotesize
    \caption{Coverage results w/ and w/o LLM/Reuse Modules. }
    \label{tab:ablation}
    \begin{adjustbox}{max width=\linewidth}
    \begin{tabular}{l|cccc}
    \toprule
    \textbf{Tool Variant} & \textbf{Line} & \textbf{Branch} & \textbf{Method} & \textbf{Class} \\
    \midrule
    \textbf{\hd}       & \textbf{40.28\%} & \textbf{27.96\%} & \textbf{43.67\%} & \textbf{53.03\%} \\
    w/o Reuse          & 34.88\%          & 23.31\%          & 34.80\%          & 42.59\%          \\
    w/o LLM            & 32.07\%          & 21.12\%          & 34.86\%          & 39.63\%          \\
    \midrule
    \textbf{\hm}       & \textbf{42.27\%} & \textbf{29.32\%} & \textbf{46.02\%} & \textbf{54.41\%} \\
    w/o Reuse          & 37.72\%          & 25.89\%          & 41.48\%          & 53.22\%          \\
    w/o LLM            & 36.80\%          & 25.10\%          & 40.62\%          & 52.72\%          \\
    \bottomrule
    \end{tabular}
    \end{adjustbox}
    \vspace{-0.4cm}
\end{table}

\vspace{-3pt}
\subsection{RQ4: Ablation}
\label{sec:rq4}

Table~\ref{tab:ablation} reports the average code coverage across all apps under different tool variants.
The full configuration consistently achieves the highest coverage across all four metrics.
Disabling probabilistic reuse and LLM guidance reduces \hd's line coverage to 34.88\% and 32.07\%, respectively.
This degradation highlights that both LLM-driven reasoning and reuse mechanism are beneficial, and the former is the primary contributor to wide exploration.

\begin{table}[t]
    \centering
    \scriptsize
    \setlength{\tabcolsep}{1.2pt} 
    \begin{minipage}[t]{0.49\linewidth}
        \centering
        \caption{Cov. under different \(k\).}
        \label{tab:simk}
        \begin{adjustbox}{max width=\linewidth}
        \begin{tabular}{l|cccccc}
        \toprule
        \textbf{Metric / \(k\)} & \textbf{5} & \textbf{6} & \textbf{7} & \textbf{8} & \textbf{9} & \textbf{10} \\
        \midrule
        Line (\%)   & 34.42 & 35.44 & 36.80 & \textbf{42.27} & 38.55 & 37.52 \\
        Branch (\%) & 22.99 & 23.77 & 25.10 & \textbf{29.32} & 25.67 & 26.10 \\
        Method (\%) & 37.53 & 38.36 & 40.62 & \textbf{46.02} & 42.31 & 40.79 \\
        Class (\%)  & 46.34 & 47.25 & 52.72 & \textbf{54.41} & 53.75 & 49.59 \\
        \bottomrule
        \end{tabular}
        \end{adjustbox}
    \end{minipage}%
    \hfill
    \begin{minipage}[t]{0.49\linewidth}
        \centering
        \caption{LLM Variants.}
        \label{tab:llm_variants}
        \begin{adjustbox}{max width=\linewidth}
        \begin{tabular}{l|ccc}
        \toprule
        \textbf{Metric} & \textbf{GPT-4o} & \textbf{GPT-3.5-turbo} & \textbf{Deepseek-R1} \\
        \midrule
        Line   & 42.27 & 41.69 & 40.31 \\
        Branch & 29.32 & 29.27 & 27.60 \\
        Method & 46.02 & 44.67 & 43.34 \\
        Class  & 54.41 & 56.06 & 55.19 \\
        \bottomrule
        \end{tabular}
        \end{adjustbox}
    \end{minipage}
    \vspace{-0.3cm}
\end{table}

\section{Discussion}
\label{sec:discussion}

\subsection{Extended Evaluation}\label{extend_eval}
We conducted two supplementary experiments to reinforce our findings.
First, for RQ1 (coverage), we evaluated \hm and \hd against \llmd~\cite{wang2025llmdroid} on its original dataset.
Strictly following the baseline's configuration (1-hour runs, 3 repetitions and method coverage), \hm achieved 10.0\% average coverage, outperforming both \hd (8.3\%) and \llmd (7.2\%)
\footnote{Detailed data is available at \url{https://github.com/hybd123/HybridDroid}}.
Second, for RQ2 (bug detection), we utilized the \themis benchmark~\cite{su2021benchmarking} (52 reproducible bugs).
\hm identified 19 bugs, surpassing \rmk (15), \fb (7), \mk (3), \gptd (1), and \au (0).
\llmd was excluded due to instrumentation incompatibilities with legacy build systems.

\vspace{-5pt}
\subsection{Sensitivity and Cost Analysis}

\noindent\textit{\textbf{Parameter Sensitivity.}}
We evaluated the tarpit detection threshold $k \in [5,10]$.
Table~\ref{tab:simk} shows that average coverage peaks at $k=8$.
Performance declines at higher values, likely due to delayed detection, while lower values tend to trigger premature interventions.
Thus, we set $k=8$ as the default, striking a balance between sensitivity and stability.
Other image-related parameters follow empirical settings; however, adaptive tuning remains a promising direction for further enhancing robustness in diverse testing environments.

\noindent\textit{\textbf{LLM Sensitivity.}}
Table~\ref{tab:llm_variants} demonstrates \hd's robustness across different LLMs. Although GPT-4o performs best, all variants (including DeepSeek-R1) surpass the baselines. This indicates our success derives from the hybrid testing paradigm itself, independent of specific LLM capabilities.


\noindent\textit{\textbf{Cost Analysis.}}
Our approach is highly cost-efficient, averaging \$0.19 per round---significantly lower than \llmd (\$0.43) and \gptd (\$9.21). Unlike \gptd's continuous querying (1,425 queries avg.), we invoke the LLM only upon detecting tarpits. This selective invocation eliminates unnecessary token consumption, making our method economically viable for large-scale testing.
\vspace{-5pt}
\subsection{Design Rationales}

\noindent\textit{\textbf{Image-Based Similarity.}}
\label{dis:image-based}
Our approach uses image-based similarity instead of layout tree comparison to detect UI tarpits.
This choice enhances robustness against the volatility of the view hierarchy.
Layout-based similarity is sensitive to minor structural changes like dynamic IDs.
Defining appropriate abstraction criteria to mitigate such structural volatility remains a significant challenge in the field~\cite{baek2016automated}.
In contrast, image-based similarity captures visual semantics and remains stable under UI fluctuations.

\noindent\textit{\textbf{Generality of our work.}}
Our LLM-guided escape mechanism is model-agnostic and is able to, in principle, enhance any automated GUI testing tools. We believe that incorporating our approach into learning-based or model-based testing tools can also yield meaningful improvements. Our results on two widely used tools (\ie, \mk and \db) serve as a proof-of-concept for this general applicability.

\vspace{-5pt}
\subsection{Threat Validity and Limitations}

\noindent \textbf{\textit{Threats to Validity.}}
A primary external threat to the validity of our work is the representativeness of the app subjects. To mitigate this, our experiments include a diverse set of open-source apps, as well as one widely used industrial app. These apps were chosen based on their popularity and relevance from GitHub and Google Play.
Furthermore, we conducted two supplementary experiments on datasets from existing works~\cite{wang2025llmdroid, su2021benchmarking} to further ensure the broad applicability and relevance of our findings to real-world scenarios. 

\noindent \textbf{\textit{Limitations and Future Work.}}
Our failure analysis (\S\ref{lesson4}) reveals that accessibility-related failures (64.56\%) stem from incomplete UI metadata.
Future work should explore multimodal approaches, such as VLMs, to recover semantics from UI components lacking labels. Refining prompts to handle complex text inputs also remains a promising direction.
Regarding scope, our approach specifically targets UI tarpits characterized by repetitive and visually similar pages.
It does not currently cover cyclic traps involving distinct UIs, such as logical loops across multiple pages.
This focus represents a deliberate trade-off, as our primary goal is to assist random testing in escaping from immediate stagnation within a specific execution path.
Consequently, our definition of UI tarpit captures a significant subset rather than the full spectrum of tarpit phenomena.
Designing a fully general-purpose tarpit detector remains a significant research challenge that we plan to explore in future work.

\vspace{-5pt}
\section{Lessons Learned}
\label{sec:lesson}

We discuss the lessons learned from our investigation.

\noindent\textbf{\textit{Lesson 1: Random testing is competitive.}}
Recent work favors sophisticated testing strategies, \eg, learning-based (like \fb) and LLM-based testing (like \llmd).
But our investigation corroborates that random testing is indeed competitive in practice~\cite{choudhary2015automated,patel2018effectiveness,wang2018empirical,mohammed2019empirical,lan2024navigating}. For example, \rmk (random testing) achieves 24.8\% branch coverage on average, surpassing \fb (23.5\%) and \llmd (21.6\%) (see \S\ref{sec:rq1_coverage}).
Our idea of using LLMs to escape tarpits during random testing is thus simple enough yet effective to be adopted in practice. 

\noindent\textbf{\textit{Lesson 2: Unleashing the power of randomness is important for bug discovery.}}
LLM is effective at helping random testing escape UI tarpits, but LLM likely suggests \emph{happy paths} (\ie, common interaction scenarios) rather than \emph{less-traveled paths} (\ie, edge-case scenarios) for UI exploration. 
Thus, our design (\ie, limiting LLM-guided UI exploration to one single step and immediately switching back to random testing when UI tarpits are escaped) aims to unleash the power of randomness for bug discovery. 
This design maximizes the chance to trigger edge cases. Indeed, \hm found many more bugs (24) than \rmk (15) and \llmd (1) (\S\ref{sec:rq2}).

\noindent\textbf{\textit{Lesson 3: GUI semantic understanding could be improved by vision-language models (VLMs).}}\label{dis:lesson3}
Our analysis (\S\ref{lesson4}) revealed that 64.56\% of escaping failures stem from incomplete or empty text labels of UI widgets when performing GUI semantic understanding. The current only text-based UI semantic understanding could be affected by low-quality text labels in practice.
Thus, we suggest that vision-language models (VLMs) could be used to improve understanding UI semantics via screenshots.

\vspace{-5pt}
\section{Related Work}
\label{sec:related-work}
\noindent\textbf{\textit{UI Exploration Tarpits in App Testing.}}
Several  studies~\cite{dong2020time,yan2020multiple,feng2023efficiency,hu2024enhancing,wang2021vet,khan2024aurora,liu2023fill,yoon2025integrating} have highlighted the UI tarpit issue in existing automated GUI testing techniques, where testing tools get stuck in some local UI regions and fail to achieve fruitful exploration. However, only \vet~\cite{wang2021vet} and \au~\cite{khan2024aurora} are explicitly designed to address this problem. \vet is the first work to explicitly define UI exploration tarpits and proposes a two-phase approach to mitigate this problem. Rather than escaping UI tarpits, \vet proposes a prevention-based approach that incurs double execution cost and may inadvertently block exploration beyond disabled states. \au, on the other hand, introduces heuristic rules to escape tarpits, but its approach is limited to eight predefined UI patterns. Other works~\cite{liu2023fill,yoon2025integrating} improve coverage by generating valid text inputs that satisfy input constraints. However, they do not target the UI tarpit problem. 
In contrast, our work focuses on a more generalized approach that dynamically identifies and mitigates UI tarpits.

\noindent\textbf{\textit{LLM-based Android GUI Testing.}}
Much work has been proposed to leverage LLMs to enhance software testing~\cite{wen2024autodroid,wang2024software,yang2024evaluation,jiang2024towards,hou2024large,lukasczyk2022pynguin,nie2023learning,guo2020audee,kongprophetagent,liu2025seeing,ju2024study,wang2024feedback,feng2024prompting,liu2024make,yoon2024intent,wang2025llmdroid,liu2023fill,cui2024large,wang2024large,gao2025llm,xue2024llm4fin,hu2025kuitest,liu2025guipilot,liu2024testing,hu2024auitestagent,yoon2025integrating}. 
Prior Android GUI testing frameworks employ LLMs to provide direct GUI interaction guidance~\cite{liu2024make}, task planning throughout the testing workflow~\cite{yoon2024intent,hu2024auitestagent}, and context-aware text input generation~\cite{liu2023fill,liu2024testing,yoon2025integrating}.
The most closely related work is \llmd~\cite{wang2025llmdroid}, which integrates LLMs with AIG tools via coverage guidance. We differ in two key aspects. 1) while \llmd aims to reduce the overhead of pure LLM-based app testing, we focus on mitigating the negative impact of UI tarpits on traditional testing to enhance efficiency. 2) \llmd uses LLMs to steer exploration toward unvisited pages based on coverage monitoring; in contrast, we monitor UI transitions to identify and escape tarpits via LLMs. Evaluation~\ref{sec:evaluation} shows that our approach outperforms \llmd in both coverage and bug discovery.




\noindent\textbf{\textit{Traditional Android GUI Testing.}}
Several studies have explored input generation for Android GUI Testing. Random-based solutions~\cite{monkey,ye2013droidfuzzer,choudhary2015automated,patel2018effectiveness,wang2018empirical,behrang2020seven,machiry2013dynodroid,sasnauskas2014intent}, focus on high-speed event injection but are semantics-oblivious.
Model-based approaches~\cite{dias2007survey,shafique2010systematic,su2017guided,mirzaei2016reducing,yang2013grey,wang2020combodroid,mao2016sapienz,yang2018static,takala2011experiences,amalfitano2012using} struggle with state abstraction and scalability. Learning-based techniques~\cite{lv2022fastbot2, pan2020reinforcement, li2019humanoid, romdhana2022deep,yasin2021droidbotx,lan2024deeply} utilize reinforcement or deep learning to predict effective actions, but remain limited by unforeseen states or long-sequence dependencies~\cite{kong2018automated,rubinov2018we,su2021benchmarking}.
Despite these advancements, existing methodologies lack specific mechanisms to identify or escape UI tarpits~\cite{wang2021vet}, frequently leading to exploration stagnation and missed bugs. Our work fills this gap by providing a targeted approach to mitigate such issues.

\vspace{-5pt}
\section{Conclusion}

UI tarpits hinders the exploration when testing GUI apps. This paper introduces a hybrid testing approach that augments random Android GUI testing with LLM, in an aim to escape tarpits. 
The evaluation results show that it significantly improves code coverage and bug detection on real-world apps.


\vspace{-5pt}
\section{Data Availability}
We have open-sourced our tools and dataset to facilitate replication and future research at~\url{https://doi.org/10.6084/m9.figshare.31861816}.

\bibliographystyle{ACM-Reference-Format}
\bibliography{citation.bib}

\end{document}